\documentclass[aps,
		prd,
		reprint,
		onecolumn,
		superscriptaddress,
		shortbibliography,
		nofootinbib,
		floatfix,
		notitlepage
		]{revtex4-1}

\usepackage{amsmath,amssymb,amsfonts}

\usepackage{natbib}
\usepackage[T1]{fontenc}
\usepackage[latin1]{inputenc}

\usepackage[dvipsnames]{xcolor}

\usepackage{hyperref}
\hypersetup{colorlinks=true,citecolor=Cyan,urlcolor=Red}

\usepackage{mathrsfs}
\usepackage{bbm}
\usepackage{slashed}
\DeclareSymbolFont{rsfs}{U}{rsfs}{m}{n}
\DeclareSymbolFontAlphabet{\mathrsfs}{rsfs}
\usepackage[mathscr]{eucal}	
\usepackage[normalem]{ulem}
\usepackage{verbatim}

\usepackage{graphicx}

\newcommand{\be}{\begin{equation}}
\newcommand{\ee}{\end{equation}}
\newcommand{\bi}{\begin{itemize}}
\newcommand{\ei}{\end{itemize}}
\newcommand{\bea}{\begin{eqnarray}}
\newcommand{\eea}{\end{eqnarray}}
\newcommand{\ud}{\mathrm{d}}		

\newcommand{\LCp}{{\scriptscriptstyle +}}

\begin{document}
%


\title{Note on the conjectured breakdown of QED perturbation theory in strong fields}
\author{A. Ilderton}
\email{anton.ilderton@plymouth.ac.uk}
\affiliation{Centre for Mathematical Sciences, University of Plymouth, Plymouth, PL4 8AA, UK}
\begin{abstract}	
Strong background fields require a non-perturbative treatment, which is afforded in QED by the Furry expansion of scattering amplitudes. It has been conjectured that this expansion breaks down for sufficiently strong fields, based on the asymptotic growth of loop corrections with increasing ``quantum nonlinearity'', essentially the product of field strength and particle energy.  However, calculations to date have assumed that the background is constant.  We show here, using general plane waves of finite duration, that observables at high quantum nonlinearity scale differently depending on whether intensity or energy is large. We find that, at high energy, loop contributions to observables tend to fall with increasing quantum nonlinearity, rather than grow.
\end{abstract}

\maketitle
\section{Introduction}
A strong electromagnetic field is characterised by a dimensionless coupling to matter which is larger than one. Hence the interaction of the field with matter cannot be treated in perturbation theory. For QED processes in a strong field the required semi-non-perturbative treatment is given by the Furry expansion of scattering amplitudes. Here the coupling between matter and generated/absorbed photons, still characterised by the fine-structure constant~$\alpha$, is treated in perturbation theory as normal, while the coupling to the strong field is treated exactly. This amounts to replacing the fermion propagator by a dressed propagator, while all other \textit{position space} Feynman rules are unchanged~\cite{Furry:1951zz}.

It has however been conjectured that for sufficiently strong fields, this semi-perturbative expansion also breaks down~\cite{Ritus1,Naroz1,Naroz2,Moroz1,Naroz3}, and the theory becomes ``fully nonperturbative''. This conjecture is based on the identification of an effective coupling parameter $\alpha\chi^{2/3}$, where $\chi$  is essentially the product of field strength and probe particle energy (see below). As field strength increases, loop corrections appear to grow with higher powers of the effective coupling, and so when $\alpha \chi^{2/3}\sim 1$ the Furry expansion breaks down. The physics of the regime $\alpha \chi^{2/3}\sim 1$ is thus unknown, but for experimental proposals for how to approach  it  see~\cite{Yakimenko:2018kih,Blackburn:2018tsn,Baumann:2018ovl}. For a review of the conjecture see~\cite{Fedotov:2016afw}. (For effective expansion parameters in relation to muon $g-2$ see~\cite{Terazawa:1968mx,Lautrup:1974ic,Miller:2007kk}.) \enlargethispage{20pt}

The calculations behind the conjecture have, however, been performed in constant crossed fields; these are the zero frequency limit of plane waves, commonly used as a first model of laser fields at ultra-high intensity~\cite{RitusRev,DiPiazza:2011tq,FedRev,King:2015tba,Seipt:2017ckc,Hartin:2018egj}. (We will typically refer to field intensity, rather than strength.) Notably, the power of $2/3$ which appears is tied to the Airy functions particular to the constant field case. This prompts the question of what happens in more general fields. Furthermore, the literature to date has tended to focus on (loop corrections to) the polarisation and mass operators, neither of which are observables. Indeed, the one-loop vertex correction has been argued to scale asymptotically as either $\alpha \chi^{2/3}$ or just $\alpha$, depending on the gauge used~\cite{Naroz3,Gusynin:1999pq}.

There are, then, several issues to address. First, we should only consider gauge invariant observables. Second, fields which decay at infinity, rather than constant fields, should be the generic case; here we will work with plane wave pulses of finite duration and arbitrary intensity, which can be treated exactly in the Furry expansion~\cite{RitusRev,DiPiazza:2011tq,King:2015tba,Seipt:2017ckc}. Third, observables do not in general depend on~$\chi$ alone, but on intensity and energy individually. A dependence solely on $\chi$ is again particular to the constant field limit, but~$\chi$ can also be made large by making the energy large. (Literature investigations of one-loop processes in pulsed fields already contain hints that the scalings attributed to the constant field case may not be universal, see for example~\cite{Dinu:2013gaa,Gies:2014jia}.) Fourth, while loop corrections have been calculated in the high intensity approximation, emissivity corrections and inclusive observables have not been discussed, and we expect higher numbers of emissions to also become more important at high intensity. In this note we will present an initial investigation of these issues using the few one-loop diagrams which have been calculated exactly in pulsed plane waves, rather than constant fields. We will contrast two high-$\chi$ limits, reached by high intensity or high energy.

We introduce relevant parameters in Sect.~\ref{SECT:PARAMS}. In Sect.~\ref{SECT:NLC} we consider nonlinear Compton scattering at tree level, and forward scattering at one loop. In Sect.~\ref{SECT:HEL} we consider photon helicity flip at one loop. In both cases we will show that observables scale differently in the two high-$\chi$ limits, demonstrating that there is no universal high-$\chi$ behaviour. The high-$\chi$ (energy) limit, in particular, does not seem to suffer from a potential breakdown of perturbation theory as the high-$\chi$ (intensity) limit does. A further consequence of our results is that, as we will show, currently used high-intensity approximations in numerical laser-plasma models are unable to properly reproduce high-energy, quantum, behaviour.  We conclude in Sect.~\ref{SECT:CONCS}. Closely related calculations have very recently appeared in~\cite{Podszus:2018hnz}, and we will find agreement with the conclusions presented there.

\section{Parameters \& invariants}\label{SECT:PARAMS}
Consider the interaction of some particle, momentum $p_\mu$, with a strong background field ${F}_{\mu\nu}(x)$. A key parameter is the ``quantum nonlinearity'' $\chi$ of the particle, defined by~\cite{RitusRev}
\be\label{chi1}
	\chi = \frac{e}{m^3}\sqrt{p^\mu F_{\mu\nu}F^{\nu\sigma}p_\sigma} \;, 
\ee
in which $e$ and $m$ are, by convention, the \textit{electron} charge and mass. We will only deal with electrons and photons here. It is clear that $\chi$ looks, in general, like the product of particle energy and field strength, with the proportionality factor depending on field- and collision geometry.  (For an electron, $\chi$ is equal to the ratio of the electric field in the rest frame, to the Schwinger field $E_S = m^2/e$.)  Since $\chi$ is a composite parameter it can be made large by increasing the particle energy, the field strength, or both. We emphasise that the physics differs depending on whether a particular value of $\chi$ is reached by going to high energy, or high intensity~\cite{Dinu:2015aci}.

Our focus here is on plane waves. For a wave characterised by the lightlike propagation direction $n_\mu$ we have
\be\label{chi2}
	\chi = \frac{n.p}{m}\frac{|{\bf E}|}{E_S} \;.
\ee
The plane wave will have some typical scale $\omega$ associated with it, be it a central frequency, inverse width, etc, and it is convenient to define dimensionless variables in terms of this scale\footnote{The physics is independent of this \textit{choice}, but not of the physical scale $\omega$. If, as for constant fields, no such scale is apparent, one can simply use the electron rest mass. Any natural scale will emerge during the evaluation of the spacetime integrals to be introduced below.}.  So, let $k_\mu = \omega n_\mu$ be a typical momentum vector associated to the wave, and $\phi = k.x$ the dimensionless ``lightfront time'' on which the wave depends. This allows us to define the dimensionless energy, $b$, and field intensity parameter, $a_0$, by
\be
	b := \frac{k.p}{m^2} \;, \qquad 
	a_0 := \frac{\chi}{b} = \frac{|e{\bf E}|}{m\omega}  \;.
\ee
(Note that $\chi = a_0 b$ is independent of the chosen scale $\omega$.) The intensity parameter $a_0$ is the coupling between the plane wave and matter, and easily exceeds unity in modern laser experiments~\cite{Sarri:2014gea,Cole:2017zca,Poder:2018ifi}. For, say, a head-on collision between an electron and the wave we have $b \simeq 2(\omega/m)\gamma$ at high energy.

Below we will compare the behaviour of processes at high $\chi$ reached via high intensity or high energy, meaning large $a_0$ or large $b$ respectively. There are many classic examples of processes which exhibit a very different dependence on a single (usually energy) invariant, compare for example the low~\cite{Euler:1935zz} and high~\cite{Akhiezer:1936vzu} energy behaviours of the photon-photon cross-section, see also~\cite{Scharnhorst:2017wzh} for a review. Here, though, we will examine how the two independent invariants $a_0$ and $b$ essentially \textit{compete} to affect different behaviours of processes in strong fields.

\section{Forward scattering and nonlinear Compton scattering}\label{SECT:NLC}
\begin{figure}[t!]
\includegraphics[width=0.6\textwidth]{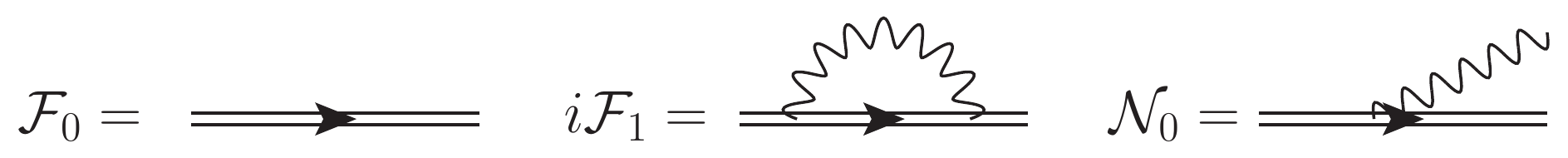}
\caption{\label{FIG:NLC} $\mathcal{F}_0$ is the Furry picture propagator in a background field, $\mathcal{F}_1$ its one-loop correction, and $\mathcal{N}_0$ tree level photon emission,  ``nonlinear Compton scattering''.}
\end{figure}

%
We begin with the diagrams in Fig.~\ref{FIG:NLC}. These are the tree level fermion propagator, its one-loop correction, and the tree level vertex, in the Furry picture. The loop diagram $\mathcal{F}_1$ (associated with the mass operator in the literature on $\alpha\chi^{2/3}$) contributes to forward scattering of the electron, and to electron spin-flip, while the vertex yields e.g.~tree level photon emission from a field-accelerated electron, or ``nonlinear Compton scattering" (NLC)~\cite{Nikishov:1964zza,RitusRev,Boca:2009zz,Heinzl:2009nd,Seipt:2010ya,Mackenroth:2010jk}. 

Consider the electron forward scattering amplitude; this is degenerate with soft emission~\cite{Yennie:1961ad,Lavelle:2005bt,Lavelle:2010hq}, so we should consider not just exclusive but also inclusive processes. Let $\mathbb{P}(b\big| a)$ be the probability of an \textit{exclusive} scattering process $a\to b$ as calculated in the Furry expansion, and let $\mathbb{P}(\sphericalangle b\big| a)$ be the inclusive probability. Then the probability of \textit{observing} an electron scattered with no photon emission above the detector resolution $\varepsilon_\text{min}$ of the system is, writing $e$ for electron and $\gamma_s$ for a soft photon, 
%
%
\be\begin{split}
	\mathbb{P}( \sphericalangle  e\big| e) = \mathbb{P}(e \big| e) + \mathbb{P}(e,\gamma_s\big|e) 
	+ \ldots &= \int\limits_{\varepsilon_\text{IR}}  | \mathcal{F}_0 + i\mathcal{F}_1 + \ldots |^2 + \int\limits^{\varepsilon_\text{min}}_{\varepsilon_\text{IR}} | \mathcal{N}_0 + \ldots |^2 + \ldots \\
	&= 1 -\int_{\varepsilon_\text{IR}} 2 \text{Im} (\mathcal F^{\star}_{0} {\mathcal F}_1) + \int\limits^{\varepsilon_\text{min}}_{\varepsilon_\text{IR}} | \mathcal{N}_0 |^2+ \mathcal{O}(\alpha^2) \;, \\
\end{split}
\ee
where $\varepsilon_\text{IR}$ is the infra-red (IR)~cutoff. By the optical theorem the one-loop imaginary part is exactly equal to the tree level probability of NLC. (For the explicit demonstration of this in plane waves see~\cite{Ilderton:2013dba}.) Hence we have
\be\begin{split}
	\mathbb{P}(\sphericalangle  e\big| e) &= 1 -\int^\infty_{\varepsilon_\text{IR}} | \mathcal{N}_0 |^2  + \int\limits^{\varepsilon_\text{min}}_{\varepsilon_\text{IR}} | \mathcal{N}_0 |^2+ \mathcal{O}(\alpha^2)  = 1 - \int\limits^\infty_{\varepsilon_\text{min}} | \mathcal{N}_0 |^2 + \mathcal{O}(\alpha^2)\;.
\end{split}
\ee
For plane waves we can take $\varepsilon_\text{min}\to 0$ without introducing any IR divergence, assuming as usual that there is no DC mode~\cite{Ilderton:2012qe}. The forward scattering probability is then equal to one minus the tree-level NLC probability, demonstrating the standard IR result that the proper inclusion of higher loop and emissivity corrections can reduce probabilities~\cite{Yennie:1961ad}. The diagrams necessary to study this in depth in the context of $\alpha \chi^{2/3}$ have not yet been calculated, however. (We return to this in the conclusions.) Here, we observe that the tree-level NLC probability also determines the one-loop forward-scattering probability, and so we turn to the explicit calculation of NLC in two, distinct, high-$\chi$ limits. 

Consider an electron, momentum $p_\mu$, colliding with a plane wave depending on lightfront time $k.x$, so $b=k.p/m^2$. The wave may be described by a potential $eA_\mu/m = a_\mu(k.x)$ which is transverse, $k.a(k.x)=0$, and where the profile $a_\mu(k.x)$ is characterised by amplitude $a_0$, with $a_0$ the intensity parameter~\cite{DiPiazza:2011tq,Seipt:2017ckc}. The electron emits a photon of momentum $k'_\mu$ and scatters. The total probability of this NLC scattering at tree level is given by an integral over the lightfront momentum fraction $s:=k.k'/k.p$ of the emitted photon, and over two lightfront times $\theta$ and $\phi$ as~\cite{Dinu:2013hsd}
\begin{align}\label{P-KOR}
\mathbb{P}( e,\gamma\big|e)	= -\frac{\alpha}{\pi b} \int\limits_0^1\!\ud s \!\int\! \ud\phi\! \int\limits_0^\infty\!\ud\theta \: \sin \bigg( \frac{s\, \theta\mu}{2b(1-s)} \bigg)
	\bigg[ \frac{1}{\mu}\frac{\partial\mu}{\partial\theta} +  \bigg(\frac{1}{2} + \frac{1}{4}\frac{s^2}{1-s}\bigg) \langle   a' \rangle^2 \theta \bigg] \;,
\end{align}
where we define, here and throughout, a floating average $\langle \cdot \rangle$ and Kibble's (normalised) effective mass $\mu$ by~\cite{Kibble:1975vz,Harvey:2012ie}
\be\label{av}
	\langle f \rangle = \frac{1}{\theta} \int\limits_{\phi-\theta/2}^{\phi+\theta/2} \!\ud (k.x)\, f(k.x) \;,
	\qquad
	\mu(\phi,\theta)= 1 - \langle  a^2 \rangle + \langle  a\rangle^2 \,.
\ee
Note that the probability (\ref{P-KOR}) is a function of $a_0$ and $b$ individually. The $s$-integral can be performed analytically~\cite{Dinu:2013hsd}.

\subsection{High intensity}
The high-intensity limit is reached when $a_0\gg1$. An old result is that the ``formation length'' of quantum processes scales in this limit as $1/a_0$, so that amplitudes can be calculated in the locally constant field approximation (``LCFA")~\cite{RitusRev}. (It is now known that this argument fails for the emission of photons with low energy~\cite{Dinu:2012tj} or low lightfront energy~\cite{Harvey:2014qla,DiPiazza:2017raw,Ilderton:2018nws}.) The scaling of processes in constant fields, on which the $\alpha \chi^{2/3}$ conjecture is based, is shared by the LCFA. We consider the slightly more general case of the LCFA because of its use in particle-in-cell codes, used to plan and analyse intense laser experiments, for a review see~\cite{Gonoskov:2014mda}. Taking $a_0\gg 1$, the LCFA for NLC is~\cite{RitusRev,DiPiazza:2017raw,Ilderton:2018nws},
\be
	\mathbb{P}( e,\gamma\big|e) \simeq -\frac{\alpha}{b}\int\!\ud\phi\!  \int\limits_0^1\! \ud s\,  \text{Ai}_1(z) + \bigg(\frac{2}{z}+\chi_\gamma\sqrt{z}\bigg)\text{Ai}'(z) \;, \quad\text{where}\quad z(\phi) := \bigg(\frac{1}{\chi_e(\phi)}\frac{s}{1-s}\bigg)^{\tfrac{2}{3}} \;.
\ee
Note that, aside from the $1/b$ prefactor, the LCFA (and constant field) expressions now depend only on $\chi$. The asymptotic scaling of the probability is obtained by expanding the Airy functions, which yields 
\be\label{NLC-HOG-I}
	\mathbb{P}( e,\gamma\big|e) \sim \frac{\alpha}{b} \int\!\ud \phi \, \chi^{2/3}(\phi) \;. 
\ee
We are only interested in typical scalings, so ``$\sim$'' indicates throughout that we neglect purely numerical factors, keeping track only of dependence on the important parameters $b$, $a_0$, $\chi$ and, below, pulse shape effects.  For pulses, the integral in (\ref{NLC-HOG-I})  generates a finite factor which, if we consider a short pulse characterised only by some width $\tau$ and use the choice of scale described in Sect.~\ref{SECT:PARAMS}, is independent of parameters. For the constant field limit proper, $\chi$ is constant and the $\phi$-integral generates a (dimensionless) length factor. Either way we have, for some peak value of $\chi$,
\be\label{NLC-HOG-I2}
	\mathbb{P}( e,\gamma\big|e) \sim \frac{\alpha \chi^{2/3}}{b}\sim \frac{\alpha \chi^{2/3}}{\gamma} (m \tau) \;. 
\ee
The first expression gives the literature scaling. The pulse length is made explicit in the second expression. As $\chi$ increases with increasing intensity, i.e.~at fixed $b$, the probability clearly grows. It easily exceeds unity, even for short pulses, demonstrating the need for higher order corrections~\cite{Dinu:2013hsd}.
%
\begin{figure}[t!]
	\includegraphics[width=0.4\textwidth]{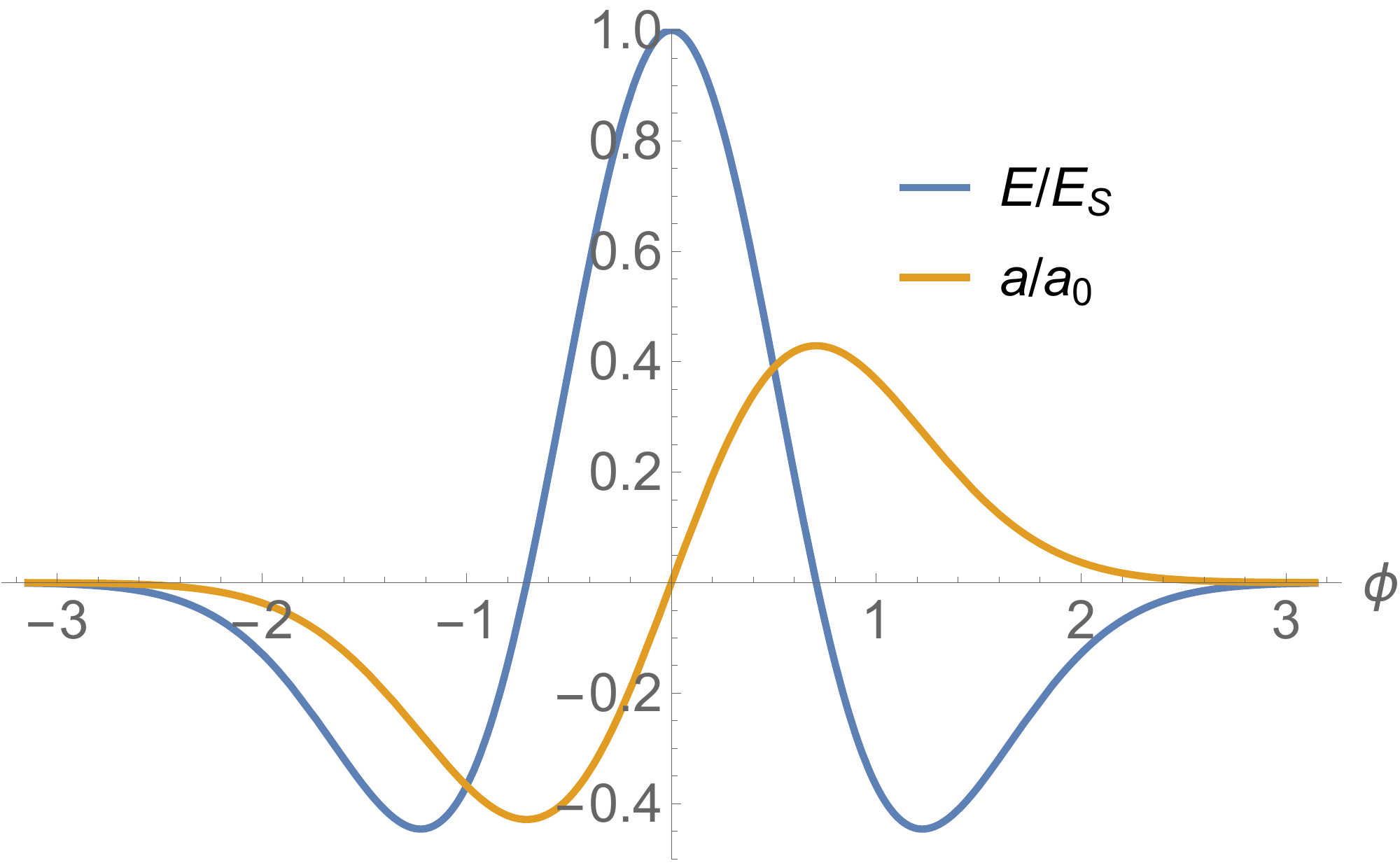}
	\caption{\label{FIG:EXEMPEL} The potential and electric field of the short pulse $a(\phi) = a_0 \, \phi \exp(-\phi^2)$, where $\phi = x^\LCp/\tau$ for some width $\tau$, as used in the text.}
\end{figure}

\subsection{High energy}
%
The high-energy, $b\to \infty$, limit of NLC may not be so well known, so we give a little more detail. Starting from (\ref{P-KOR}), we rescale $\theta \to b \theta$. Then we use the assumption that the plane wave has finite duration, for then $\mu(\phi,\infty) =1$~\cite{Hebenstreit:2010cc,Harvey:2012ie}, i.e.~the effective mass at large distance equals the free mass. To leading order in $b\gg 1$ we can then replace $\mu\to 1$. It may be checked that the first term in square brackets of (\ref{P-KOR}) only contributes sub-leading terms in the high-energy limit, so we turn to the second term. We Fourier transform each factor of $a'$ appearing. The overall $\ud\phi$ integral is then trivial, setting the Fourier variables to be equal and opposite, such that we obtain $ {\tilde a(\nu)}.\tilde{a}^\star(\nu)$ in the integrand for $\nu$ the remaining Fourier variable. Writing $t\equiv s/(1-s)$, the $\theta$-integral reduces at this stage to
\be
	4\int\limits_0^\infty\!\frac{\ud\theta}{\theta} \sin^2 \frac{\nu \theta}{2} \sin \frac{\theta t}{2b} = \pi\, \Theta(2b |\nu| - t) \;.
\ee
This puts a limit on the $s$--integral, which is most easily performed by changing variables to $t$. The final result for the leading order high energy limit is\footnote{Interestingly, this method is similar to that needed to obtain the $s\to 0$ (\textit{low} lightfront energy) limit of the differential probability. While this appears to be zero from (\ref{P-KOR}), the correct nonzero value is only obtained after performing the lightfront time integrals~\cite{DiPiazza:2017raw,Blackburn:2018sfn}.}
\be\label{Fourier}
	\mathbb{P}( e,\gamma\big|e)  = -\frac{\alpha}{4b}\int\!\frac{\ud \nu}{2\pi}\, {\tilde a(\nu)}.\tilde{a}^\star(\nu)\bigg(\frac{1}{2} + \log 2 b |\nu|\bigg) + \ldots \;.
\ee
This describes the convolution of the plane wave intensity profile ${\tilde a(\nu)}.\tilde{a}^\star(\nu)$ with the term in large brackets, which is just the high-energy limit of \textit{ordinary} Compton scattering of an electron, momentum $p_\mu$, against a photon of momentum $\nu k_\mu$ (see e.g.~\cite[\S 6.1]{Kaku:1993ym}). The whole result is quadratic in $a_0$ and is, interestingly, exactly the same as the high-energy limit of the wholly \textit{perturbative}, small $a_0$, calculation. The remaining Fourier integral is finite for finite pulses, and it is clear that the asymptotic behaviour of the probability is
\be\label{NLC-EN-1}
	\mathbb{P}( e,\gamma\big|e) \sim  \frac{\alpha a_0^2}{b}  \log b \;.
\ee
This may be confirmed with an example. Consider the pulse in Fig.~\ref{FIG:EXEMPEL}, with $a(\phi) = a_0 \phi \exp(-\phi^2)$ and linear polarisation. Then we can evaluate the Fourier integral in (\ref{Fourier}) exactly to find
\be
	\mathbb{P}( e,\gamma\big|e) = \frac{\alpha a_0^2}{b} \frac{1}{16}\sqrt{\frac\pi2}  \big(\log b + \frac{1}{2}\log 2 + \frac{3}{2}- \frac{\gamma_E}{2}\big) + \ldots \;.
\ee

\subsection{Discussion}
The expressions (\ref{NLC-HOG-I2}) and (\ref{NLC-EN-1}) give two different high-$\chi$ limits of both NLC at tree level, and electron forward scattering at one loop. These two limits have different functional forms, and a comparison is facilitated by making it explicit that the high intensity limit corresponds to increasing $\chi$ by increasing $a_0$, at fixed $b$, while the high energy limit corresponds to increasing $\chi$ by increasing $b$ at fixed $a_0$. This allows us to write (\ref{NLC-HOG-I}) and (\ref{NLC-EN-1}) as 
\be\label{NLC-EN-2}
	\underset{\chi\sim a_0\to\infty}{\mathbb{P}( e,\gamma\big|e)} \sim \frac{\alpha \chi^{2/3}}{b} \;, \qquad \qquad
	\underset{\chi\sim b\to\infty}{\mathbb{P}( e,\gamma\big|e)} \sim \alpha \frac{a_0^3}{\chi} \log \chi \;. 
\ee
Hence the nonlinear Compton probability \textit{falls} with increasing $\chi$ at high energy, rather than increasing as it does at high intensity. Furthermore, this implies that at high-$\chi$ reached via high energy, the loop correction to forward-scattering is \textit{smaller} than the tree level contribution, again in contrast to the high intensity limit. This shows explicitly that there is no universal high-$\chi$ behaviour of observables. Also, at least for the observables considered here, the increase in size of higher-loop corrections at high intensity does not appear at high energy.

The LCFA is a commonly employed tool which allows for progress where analytic results are lacking, especially in the consideration of inhomogeneous background fields with realistic spacetime structure. It is already known that the CCF and LCFA approximations, as currently employed, give incorrect results at low photon energy~\cite{Dinu:2012tj} and low lightfront momentum~~\cite{Harvey:2014qla,DiPiazza:2017raw,Ilderton:2018nws}, and we can now show that they also fail in the high-energy regime.

The LCFA for nonlinear Compton is (aside from the $1/b$ prefactor) a function of $\chi$ alone, and the high-$\chi$ limit is therefore (\ref{NLC-HOG-I}); it is impossible to obtain the correct high energy behaviour in (\ref{NLC-EN-2}) from the LCFA expression. The scaling is different, and furthermore the logarithmic dependence in (\ref{NLC-EN-2}) is missed. The reason for this is that the LCFA requires, at the level of the integrated rates as considered here, not just $a_0\gg 1$ but $a^2_0\gg b$~\cite{khok,Dinu:2015aci}, so that making the assumption that the LCFA holds \textit{precludes} the possibility of going to high-energy. Our example of NLC makes this concrete; the high-intensity limit (\ref{NLC-HOG-I2}) comes from the \textit{first} term in square brackets of (\ref{P-KOR}), while the high-energy limit (\ref{NLC-EN-1}) comes from the \textit{second} term.  This shows straightforwardly that the two limits do not commute. 

It follows that approximations used in particle-in-cell codes break down at high energy. It is therefore important to understand, when modelling potential experiments, whether high $\chi$ is being reached via high energy or high intensity.

\section{Helicity flip}\label{SECT:HEL}
\begin{figure}[t!]
\includegraphics[width=5cm]{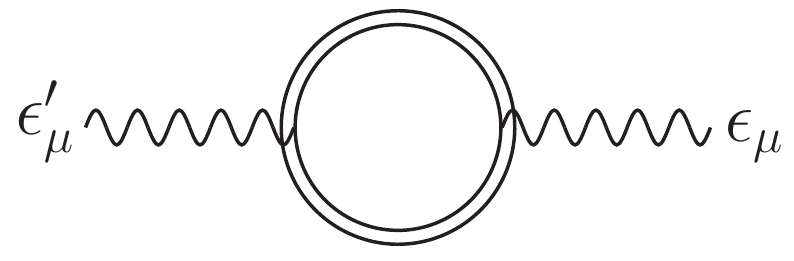}
\caption{\label{FIG:FLIP} The one-loop diagram contributing to helicity flip and forward scattering at one loop (and, via the optical theorem, pair production at tree level).}

\end{figure}
For our next example we consider photon helicity flip. In a plane wave a single photon can produce pairs, or otherwise only scatter forward (due to the many symmetries of plane waves), and may flip helicity due to loop interactions with the wave.  Let the photon have momentum $k'_\mu$, so that $b=k.k'/m^2$ is the energy parameter, and let $\epsilon_\mu$ and $\epsilon'_\mu$ be two orthogonal helicity polarisation vectors, see Fig.~\ref{FIG:FLIP}. Then the probability $\mathbb{P}(\gamma'|\gamma)$ of helicity flip is given at one-loop by $\mathbb{P}(\gamma'|\gamma) = |T|^2$, where the amplitude $T$ may be expressed, using lightfront field theory~\cite{Brodsky:1997de,Heinzl:2000ht,Bakker:2013cea}, as a double integral~\cite{Dinu:2013gaa,Meuren:2014uia} over two lightfront times $\phi$ and $\theta$. Explicitly~\cite{Dinu:2013gaa},
\be\label{T-FLIP-resultat}
\begin{split}
T= 
-\frac{\alpha}{\pi}\frac{1}{b}
\int\limits_{-\infty}^\infty\!\ud \phi \!\int\limits_0^\infty\!\ud\theta\,\theta\
  \bigg( \mathcal{K}_1\big(\tfrac{\theta \mu}{b}\big)
  \big( \bar{a}_\theta a_\theta -\tfrac{1}{4}\bar{a}_\phi a_\phi \big) 
 + \mathcal{K}_2\big(\tfrac{\theta \mu}{b}\big) 
 \big(\tfrac{1}{2}\bar{a}_\phi a_\theta-\tfrac{1}{2}\bar{a}_\theta a_\phi\big) \bigg) \;,
\end{split}
\ee
in which we have defined  
\be
	a_\phi = \partial_\phi \epsilon.\langle a \rangle \;,
	\quad \bar a_\phi = \partial_\phi \epsilon^{\prime\star}.\langle a \rangle \;,
	\quad a_\theta = \partial_\theta \epsilon.\langle a \rangle \;,
	\quad \bar a_\theta = \partial_\theta \epsilon^{\prime\star}.\langle a \rangle \;,
\ee
and the two $\mathcal{K}$-functions are combinations of modified Bessel functions arising from the integral over the virtual lightfront momentum fraction in the loop~\cite{Dinu:2013gaa}
\be\begin{split}
	\mathcal{K}_1(x) &= 
			ix e^{-ix} \big(K_1(ix)- K_0(ix)\big) 
			= \int\limits_0^1 \!\ud s\,  \exp\bigg[\frac{-ix}{2s(1-s)}\bigg] \;, \\
	\mathcal{K}_2(x) &=
		(1-i\partial_x)\mathcal{K}_1(x) \;.
\end{split}
\ee
\subsection{High intensity}
Consider a linearly polarised plane wave. (The same results are found for other polarisation choices, but the intermediate expressions are not as clear.)  Then the high intensity, LCFA, approximation to the flip amplitude is~\cite{Dinu:2013gaa}
\be\label{Tapprox}
	T \sim \frac{\alpha}{b}\int\!\ud\phi \int\limits_0^1\!\ud s\, \frac{1}{z} \big(\text{Ai}'(z) - i \text{Gi}'(z)\big) \;, \quad\text{where}\quad z= \bigg(\frac{1}{\chi_\gamma(\phi)s(1-s)}\bigg)^{2/3} \;,
\ee
and $\text{Gi}$ is the Scorer function. As before, we simply replace the $\phi$-integral by a volume factor for the constant field case. The individual dependence on $a_0$ and $b$ seen in the integrand of~(\ref{T-FLIP-resultat})~\cite{Dinu:2013gaa} is again replaced by a dependence only on their product, $\chi$. The asymptotic behaviour of (\ref{Tapprox}) is easily extracted from the known expansion of the Airy and Scorer functions as
\be\begin{split}\label{FLIPFLIP1}
	T &\sim \frac{\alpha}{b} \int\!\ud\phi \, \chi^{2/3}_\gamma(\phi) \sim \frac{\alpha \chi^{2/3}}{b} \quad \implies \quad 	\mathbb{P}(\gamma'|\gamma) \sim \frac{\alpha^2\chi^{4/3}}{b^2} \;.
	\end{split}
\ee
This is also the scaling in the constant field case~\cite{Naroz1968,Ritus1,Karbstein:2013ufa,Fedotov:2016afw} (for which the numerical value given by the $\phi$-integral again becomes a length factor).
%

\subsection{High energy}
%
We turn to the high-energy limit of helicity flip. Consider the $\mathcal{K}_1$-term in (\ref{T-FLIP-resultat}). The high energy limit may be calculated simply by expanding in powers of the small parameter $1/b$, replacing $\mathcal{K}_1 \to 1$. By writing the averages in terms of Fourier integrals it is then easy to perform the $\phi$ and $\theta$ integrals. Turning to the $\mathcal{K}_2$-term, we first integrate by parts in $\theta$ (the boundary terms are zero), such that we take the derivative of $\theta \mathcal{K}_2$. (The integration of the averages is easily performed in Fourier space.) This has the effect of improving convergence under the integral, such that we can again expand in powers of $1/b$. Expanding $-\partial_\theta(\theta \mathcal{K}_2)$ in this way the leading order terms are
\be
	\log b -2 - \gamma_E   - \log (i \theta \mu/2) + \mathcal{O}\bigg(\frac{1}{b}\bigg) + \mathcal{O}\bigg(\frac{\log b}{b}\bigg) \;.
\ee
This multiplies a function which falls like $1/\theta^2$ for large $\theta$. Hence for large enough $b$ the $\log b$ term will dominate, and we can, for pulsed fields, neglect the $\log \theta \mu$ term. (This can be confirmed by changing variables $\theta \to \theta b$ as before and expanding again; one finds only higher order terms in $1/b$.) Carrying out the integrals, the final result may be written in terms of the Fourier transform ${\tilde a}_\mu(\nu)$ as
\be\label{T11}
	T \sim \frac{\alpha}{\pi b}\int\! \frac{\ud \nu}{2\pi}\, {\bar\epsilon}'.{\tilde a}(\nu)\,{\epsilon}.{\tilde a}^\star(\nu) \bigg(\frac{1}{2} - \frac{i\pi}{2}\text{sign}(\nu) \log b\bigg) + \ldots
\ee
For generic pulses the Fourier integral converges. The first term in (\ref{T11}) is nonzero for helicity states, but whether the second term survive depends on the polarisation of the field. Generically, then, for a helicity- and polarisation--dependent constant $\sigma$ we have, 
\be\label{FLIPFLIP2}
	T \sim \frac{\alpha a_0^2}{b} (1 + \frac{\sigma}{2}\log b) \implies 
	 \mathbb{P}(\gamma'|\gamma) \sim \frac{\alpha^2a_0^4}{b^2}\big( 1 + \sigma \log b \big) \;.
\ee
\subsection{Discussion}
In order to compare the high-$\chi$ (intensity) limit (\ref{FLIPFLIP1}) of helicity flip with the high-$\chi$ (energy)  limit (\ref{FLIPFLIP2}) we again rewrite the latter in terms of $\chi$, increasing with $b$ at fixed $a_0$. This gives, to leading order,
\be
		\underset{\chi\sim a_0 \to\infty}{\mathbb{P}(\gamma'|\gamma)}   \sim \frac{\alpha^2\chi^{4/3}}{b^2} \;,
		\qquad
		\qquad
		\underset{\chi\sim b \to\infty}{\mathbb{P}(\gamma'|\gamma)} \sim \alpha^2\frac{a_0^6}{\chi^2}\big( 1+ \sigma \log \chi  \big) \;.
\ee
These two high-$\chi$ limits exhibit the same trends as for NLC, above. In the high-$\chi$ (energy) limit, the probability of helicity flip is manifestly decreasing with $\chi$, unlike in the high intensity limit. This agrees with the numerical evaluation of the exact result~(\ref{T-FLIP-resultat}) given in~\cite{Dinu:2013gaa}. The LCFA approximation again fails to capture the logarithmic behaviour at high energy, and for the same reason as above; the high intensity limit precludes taking the high energy limit. Note again that the high-$\chi$ (energy) limit coincides with that of the lowest-order perturbative calculation of the process. There are, as for the other processes considered here, higher-order corrections depending on $a_0^2/b$.

\section{Conclusions}\label{SECT:CONCS}
We have considered the behaviour of QED scattering probabilities in background plane wave fields, in the high-$\chi$ limit. This was in the context of the conjectured breakdown of (Furry picture) perturbation theory in the regime $\alpha \chi^{2/3} \sim 1$. We have shown, though, that there is no unique high $\chi$ behaviour. The high intensity and high energy limits, which both give high $\chi$, yield different scalings. At high intensity, both constant field results and locally constant field approximations of pulsed field results show the same power law scaling with $\alpha\chi^{2/3}$. The high energy limit, on the other hand, shows for general pulsed fields a logarithmic dependence on $\chi$ which is typical of QED.

Further, we have seen that observables tend to \text{fall} with $\chi$ at high energy, rather than rise, as at high intensity.  This suggests that the high-$\chi$ limit of scattering processes in strong fields, reached via high energy, may not exhibit the perturbative breakdown attributed to the high-$\chi$ limit reached via high intensity. More work is however needed to better judge this, especially since only a few diagrams have been calculated in the level of detailed needed.

We have also seen that the high-energy behaviour of the considered rates cannot be recovered once the locally constant crossed field approximation (LCFA) has been made: making that approximation on the parameter space precludes being able to take the high-energy limit. An immediate consequence is that LCFA-based particle-in-cell simulations used to model laser-matter interactions do not correctly capture high-energy quantum effects.

Note that we do not disagree with previously calculated constant field scalings. However, we have stressed that it is important to consider observables. It would in particular be interesting to examine \textit{inclusive} observables, and to see how the consistent inclusions of both loops~\cite{Naroz1,Naroz2,Moroz2} and emission affects their behaviour. It may be that this reduces probabilities, as happens with the exponentiation of infra-red corrections in QED~\cite{Yennie:1961ad} (and in background plane waves~\cite{Ilderton:2012qe}), and so brings the perturbative expansion back under control.  Alternatively, if we view the breakdown of perturbation theory as a breakdown of the background field approximation, then it may be that including some form of back-reaction~\cite{Seipt:2016fyu} at each order of perturbation theory is enough to give a better behaved series. These are challenging and interesting topics for future study.

Finally, we note that the high energy limit of the observables we have considered is equal to that obtained if the plane wave background is treated perturbatively, instead of exactly using the Furry expansion. This may offer simplifications for the calculation of more complex processes at high energy.

\acknowledgments
\noindent \textit{Thanks to Tom Heinzl and Ben King for useful discussions.} \textit{As this manuscript was being prepared, \cite{Podszus:2018hnz} appeared on the arXiv, the subject of which is very similar to our own. Encouragingly, while our methods differ, our conclusions are in agreement. Thanks to Antonino Di Piazza for correspondence on this matter.}

\end{document}